\documentclass[11pt,twoside,usenatbib]{article}

\usepackage{asp2006}
\usepackage{epsf}
\usepackage{psfig}
\usepackage{lscape}
\usepackage{natbib}
\def\yr{\hbox{${\rm\thinspace yr}$}}
\def\Msol{\hbox{${\rm\thinspace M_{\odot}}$}}
\def\Msolpyr{\hbox{${\rm\Msol\yr^{-1}\,}$}}

\begin{document}
\title{Non-thermal emission from Massive Young Stellar Objects}
\author{E.~R.~Parkin, J.~M.~Pittard, M.~G.~Hoare}   
\affil{School of Physics and Astronomy, The University of Leeds,
Woodhouse Lane, Leeds LS2 9JT, UK}  

\begin{abstract} 
In the young stellar object (YSO) phase of their lives, massive stars
drive bi-polar molecular outflows. These outflows produce beautiful,
often hourglass shaped, cavities. The central star possesses a
powerful stellar wind ($v \sim$ 2000$\;{\rm km s^{-1}}$), and possibly
a dense equatorial disk wind ($v \sim$ 400$\;{\rm km s^{-1}}$), which
collide with the inner surface of the bi-polar cavity and produces hot
($T \sim~10^{5}-10^{8}$ K) shocked plasma. A reverse shock is formed
at the point where the ram pressure between the preshock flow balances
the thermal pressure of the postshock flow and provides a site for the
acceleration of non-thermal particles to relativistic
energies. Hydrodynamical models of the wind interaction, coupled with
calculations of the non-thermal energy spectrum, are used to explore
the observable synchrotron and gamma-ray emission from these objects.

\end{abstract}



\section{Introduction}

The formation of massive stars involves outflows \citep{Garay:1999,
Reipurth:2001, Banerjee+Pudritz:2006, Banerjee:2007}. Bi-polar
cavities around YSOs are commonly observed in star formation
\citep{Garay:1999}, with collimation factors of $\sim2-10$ for massive
YSOs (MYSOs) \citep{Beuther:2002b, Davis:2004}. Numerical simulations
of high-mass star formation show that outflows are a by-product of the
collapse process which forms the massive star, and may be due to
radiation pressure and magnetic fields
\citep{Yorke:2002,Banerjee:2007}. The termination shock of an MYSO jet
can provide a site for particle acceleration \citep{Araudo:2007}.

The widths of IR recombination line emission observed from MYSOs
indicates the presence of dense outflows with velocities ranging from
100 to $>340\;{\rm km\thinspace s}^{-1}$
\citep{Drew:1993,Bunn:1995}. Further confirmation of outflows has come
from high angular resolution radio observations
\citep[e.g.][]{Hoare:1994,Hoare:2006}. One explanation would be a disk
wind generated when UV flux from the star is absorbed and re-emitted
by the material at the surface of the disk \citep[e.g.][]{Drew:1998}.

X-rays have been detected from deeply embedded MYSOs in star forming
regions by the \textit{Chandra X-ray observatory} (hereafter
\textit{Chandra})\citep[e.g.][]{Wang:2007,Wang:2009}. An initial
observation of Mon R2 detected X-ray emission from the intermediate
mass objects consistent with the hard and highly time variable X-ray
emission caused by magnetic flaring activity between the star and disk
\citep{Kohno:2002}. Further analysis of the same data, coupled with
high resolution near-IR interferometry identified the separate
constituents of Mon R2 IRS3 \citep{Preibisch:2002}. The X-ray emission
from IRS3 A and C, with a count rate of $0.166\pm 0.041\;{\rm
ks^{-1}}$ for the former, could not be explained by the standard
scenario for massive stars \citep[i.e. wind embedded shocks produced
by instabilities inherent in radiatively-driven winds -
see][]{Owocki:1988}, yet the estimated stellar masses of these objects
implies they will have radiative outer envelopes which poses problems
for the generation of X-rays through magnetic recombination between
the star and disk. In this work we examine another potential source of
X-rays, namely the collision between the outflowing stellar and disk
winds and the infalling envelope.
 
We find that the interaction of the stellar and disk wind with the
cavity wall produces shock heated plasma which emits X-rays in
agreement with \textit{Chandra} observations (Parkin et al. 2009, in
prep). The reverse shock, plus weak shocks within the hot gas, provide
a site for particle acceleration. Therefore, a timely question,
considering the recent launch of the Fermi Gamma-ray Space Telescope
(\textit{Fermi}), is whether the winds-cavity interaction produces an
observable non-thermal energy spectrum extending up to $\gamma$-ray
energies. To this aim we have developed our existing model for thermal
X-ray emission to consider non-thermal radio and $\gamma$-ray
emission.

\section{Winds-cavity interaction}

The model consists of a MYSO situated at the centre of a previously
evacuated bipolar cavity which is surrounded by infalling molecular
cloud material. We include the stellar wind and a disk wind which
emanates from the surface of the circumstellar accretion disk; both
are assumed to be at terminal velocity. Due to the spatial scales
under consideration we do not attempt to model the radiative driving
of the winds as this requires high spatial resolution in the vicinity
of the star/disk \citep[e.g.][]{Owocki:1994,Proga:1998}. For
simplicity we adopt an angle dependant wind prescription, whereby the
stellar wind occupies a region up to a polar angle of $60^{\circ}$ and
the disk wind occupies the region from $60^{\circ}-85^{\circ}$. The
density and velocity distibutions for the infalling cloud material are
described by the equations of \cite{Terebey:1984}. The morphology of
the pre-existing outflow cavity is described by a simple analytical
prescription similar to that of \cite{Alvarez:2004}.

We consider a $\sim30$\Msol O8V star with a mass-loss rate of
$10^{-7}$\Msolpyr and terminal wind speed of $2000\;{\rm km
s^{-1}}$. The disk wind has a mass-loss rate of $10^{-6}\Msolpyr$ and
a terminal wind speed of $400\;{\rm km s^{-1}}$. The unshocked winds
are assumed to be at a temperature of $10^{4}\;$K. The mass infall
rate for the cloud is $2\times10^{-4}\Msolpyr$ and the cavity opening
angle is $30^{\circ}$. The winds-cavity interaction was followed for a
simulation time of $t=2000\;$yrs.

\begin{figure}
\begin{center}
    \begin{tabular}{ll}
      \resizebox{50mm}{!}{{\includegraphics{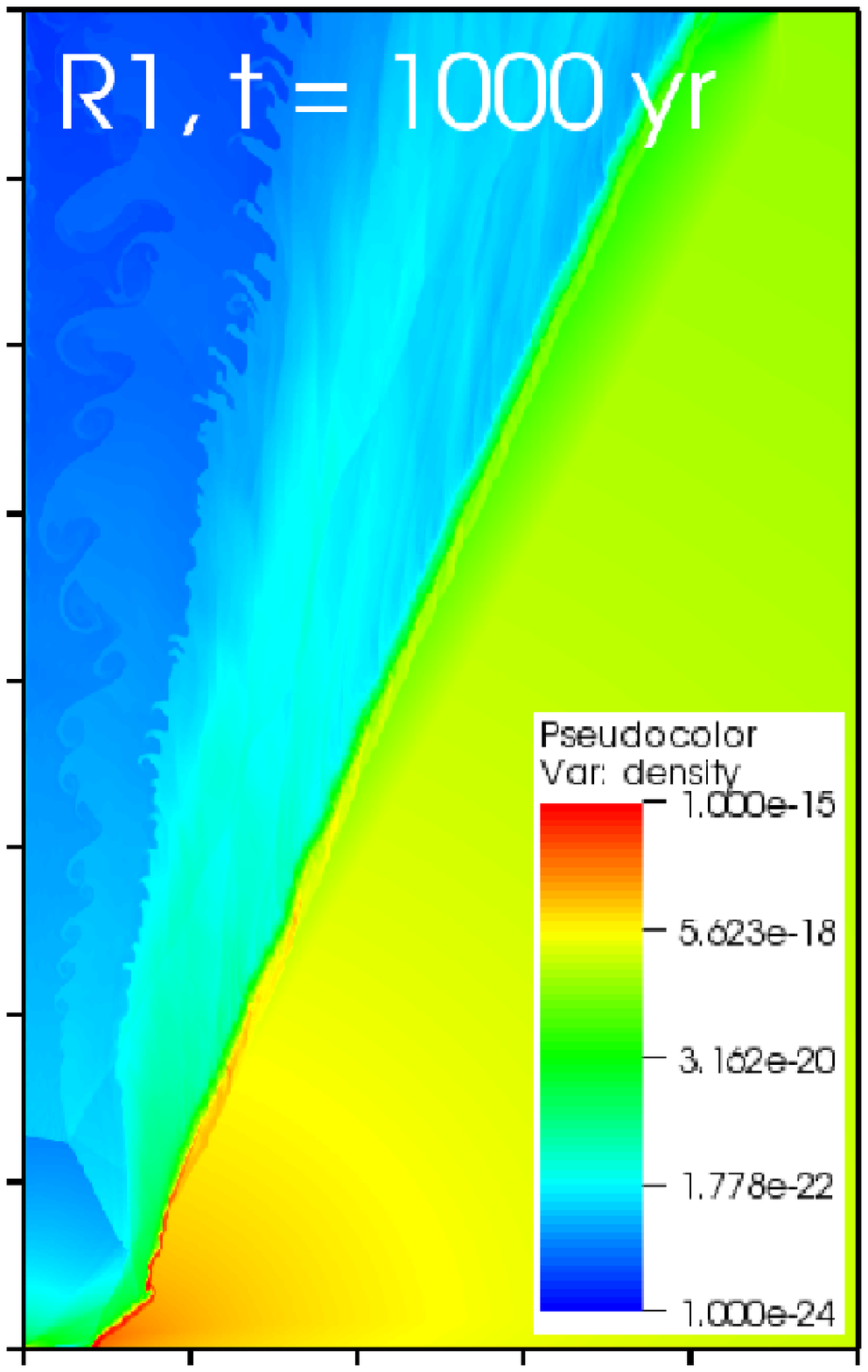}}} &
      \resizebox{50mm}{!}{{\includegraphics{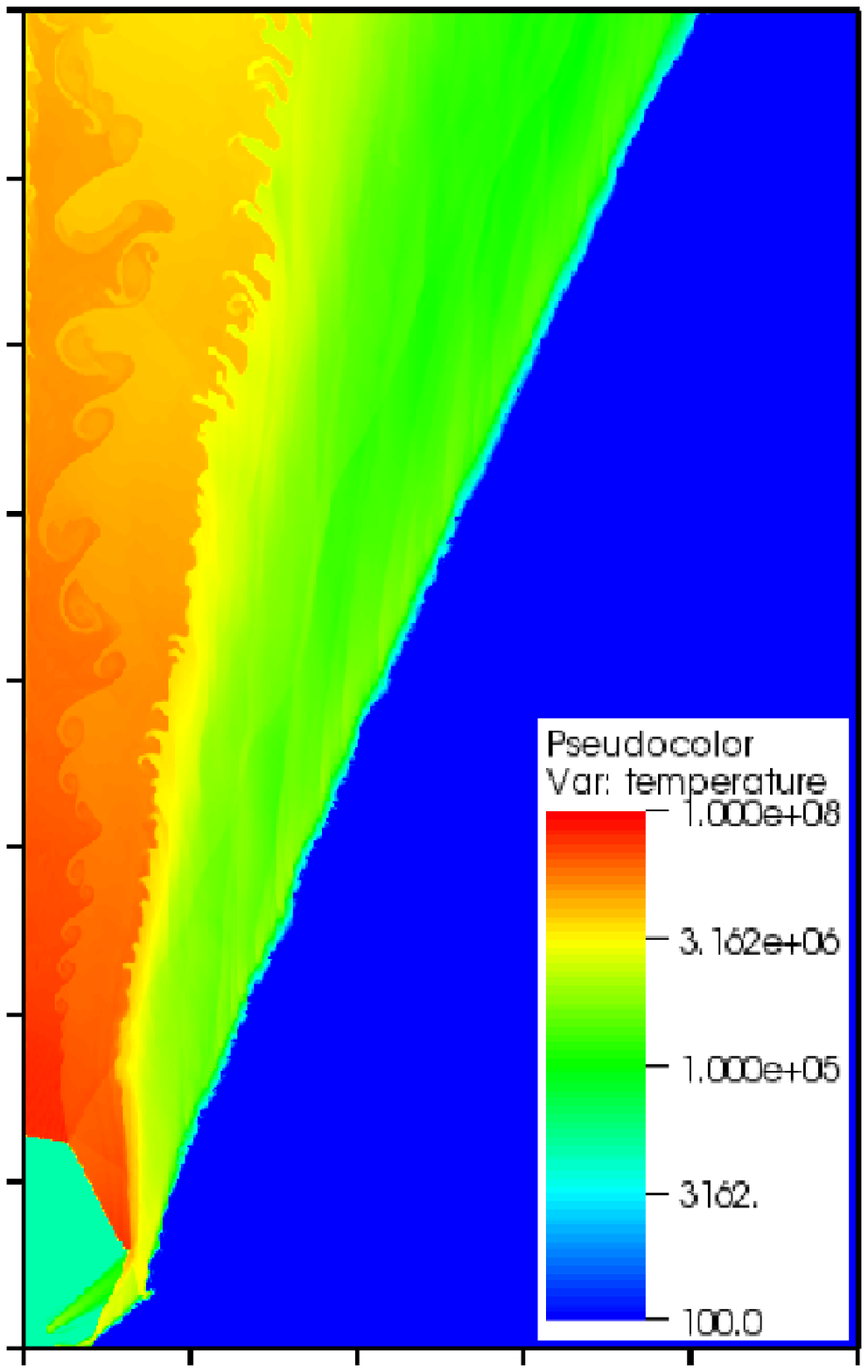}}} \\
    \end{tabular}
    \caption[]{Density (left) and temperature (right) snapshots taken
      from a hydrodynamic model of the wind-cavity interaction at
      $t=1000$ yrs. The tick marks on the axis correspond to a
      distance of $10^{16}\;$cm.}
    \label{fig:hydrosim}
\end{center}
\end{figure}

Examining Fig.~\ref{fig:hydrosim} we see that the disk wind, which has
a post-shock temperature of $\sim 10^{6}\;$K, lines the cavity wall
beyond which resides the infalling cloud material. The hotter
post-shock stellar wind fills a volume closer to the pole. A reverse
shock is established within $\sim 10^{16}$cm of the star/disk which
can be clearly seen in the temperature plot of
Fig.~\ref{fig:hydrosim}. The shape and position of the reverse shock
fluctuates with time causing variability in the observed emission on
timescales of $\sim 50\;$yrs. The shocked stellar and disk winds reach
temperatures up to $\sim 10^{8}\;$K and $\sim 10^{6}\;$K,
respectively.

\begin{figure}
\begin{center}
    \begin{tabular}{l}
      \resizebox{90mm}{!}{{\includegraphics{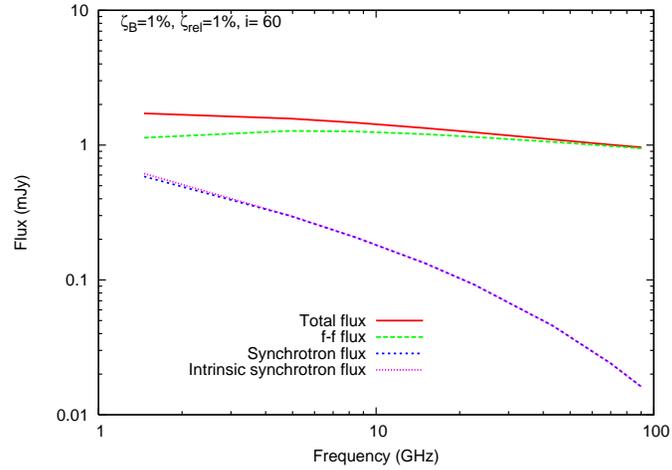}}} \\
    \end{tabular}
    \caption[]{1-100 GHz radio spectrum calculated from the
    hydrodynamic simulation shown in Fig.~\ref{fig:hydrosim} using an
    inclination angle of 60$^{\circ}$. In this plot $\zeta_{\rm
    B}=\zeta_{\rm rel}=1\%$.}
    \label{fig:radio_emission1}
\end{center}
\end{figure}

\begin{figure}
\begin{center}
    \begin{tabular}{l}
      \resizebox{90mm}{!}{{\includegraphics{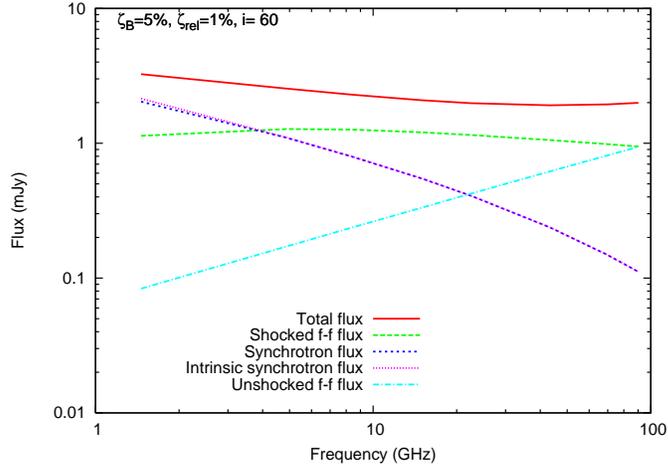}}} \\
    \end{tabular}
    \caption[]{Same as Fig.~\ref{fig:radio_emission1} except
    $\zeta_{\rm B}=5$. The contribution to the free-free flux from the
    unshocked stellar and disk winds and the shocked plasma are
    shown.}
    \label{fig:radio_emission2}
\end{center}
\end{figure}

Thermal X-ray emission in agreement with \textit{Chandra} observations
of MYSOs \citep[e.g.][]{Kohno:2002,Preibisch:2002,Giardino:2004} is
produced by the shocked gas adjacent to the reverse shock. In
particular, hard X-rays are produced ($E > 5\;$keV), the simulated
ACIS-I count rate is $\sim 0.2\;$ks$^{-1}$, and the visual extinction
to the star is $\sim30-100\;$mag (corresponding to a column of
$5.7-190\times10^{21}\;{\rm cm^{2}}$). We note that the emission
weighted column density \citep{Parkin:2008}, which is indicative of
the column density to the regions of highest intrinsic emission, is
higher than the column density to the star.

\begin{figure}
\begin{center}
    \begin{tabular}{ll}
      \resizebox{90mm}{!}{{\includegraphics{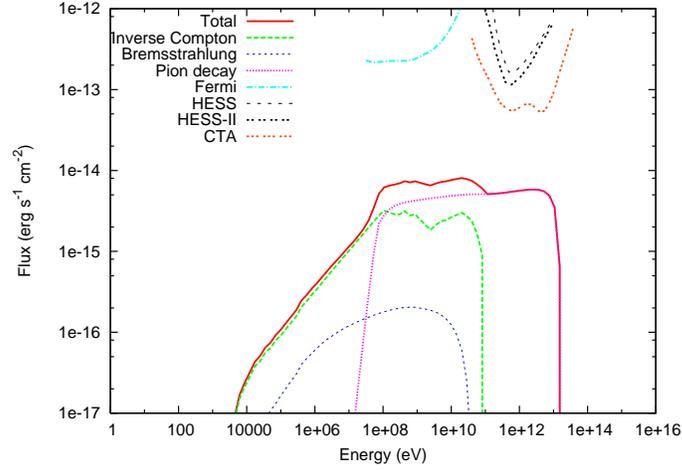}}} \\
    \end{tabular}
    \caption[]{Non-thermal energy spectrum calculated from the
    hydrodynamic simulation shown in Fig.~\ref{fig:hydrosim} where
    $\zeta_{\rm B}=\zeta_{\rm rel-e}=\zeta_{\rm rel-p}=5\%$. The
    sensitivity curve for the Cherenkov Telescope Array (\textit{CTA})
    is from \cite{Bernlohr:2008}.}
    \label{fig:gamma-ray_emission1}
\end{center}
\end{figure}

\begin{figure}
\begin{center}
    \begin{tabular}{ll}
      \resizebox{90mm}{!}{{\includegraphics{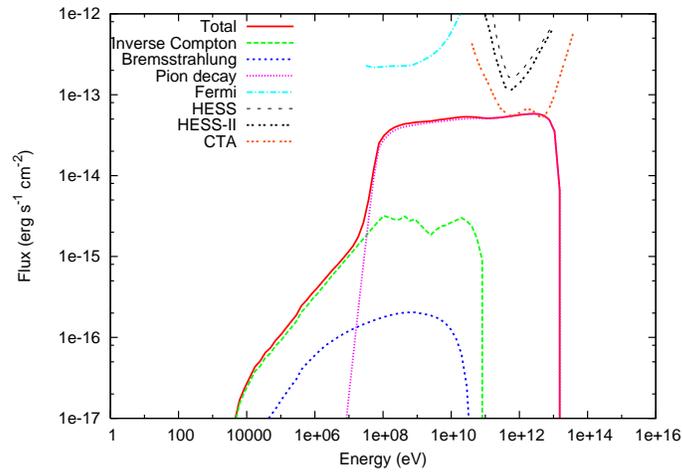}}} \\
    \end{tabular}
    \caption[]{Same as Fig.~\ref{fig:gamma-ray_emission1} except
    $\zeta_{\rm rel-p}=50\%$}
    \label{fig:gamma-ray_emission2}
\end{center}
\end{figure}

\section{Radio Emission}

The reverse shock present in the simulations provides a site for the
acceleration of non-thermal particles to relativisitic energies. We
calculate the synchrotron emission and absorption, the Razin effect,
and Inverse Compton (IC) cooling of electrons. The non-thermal and
magnetic field energy densities are assumed to be proportional to the
thermal energy density. The non-thermal particle distribution is
described by a power-law distribution with $p=2$. The ratio of the
magnetic to thermal energy density is $\zeta_{\rm B}$, and similarly
the ratio of the relativistic particle to thermal energy density is
$\zeta_{\rm rel}$. The free-free emission from the ionized winds is
also calculated.

Fig.~\ref{fig:radio_emission1} shows the resulting radio spectrum when
$\zeta_{\rm B}=\zeta_{\rm rel}=1\%$ is assumed. The total spectrum has
a flat slope, indicative of optically thin plasma. The free-free
emission from the unshocked winds is lower than the contribution from
the shocked winds inside the cavity. The thermal emission is dominant;
the synchrotron emission only has a small impact on the total spectrum
at low frequencies. Increasing the post-shock magnetic energy density
($\zeta_{\rm B}=5\%$) causes a rise in the magnitude of the
synchrotron emission (Fig.~\ref{fig:radio_emission2}) and alters the
slope of the oberved spectrum. With this in mind the observed radio
spectrum could be used to constrain parameter values for the magnetic
and relativistic particle energy density. However, a major difficulty
with this approach lies in attaining observational constraints due to
the low surface brightness and large spatial extent of the emission
region; typical observations of massive star forming regions strive
for high spatial resolution which resolves out emission on the scales
we are interested in.

\section{Gamma-ray emission}

To examine the possibility of observable $\gamma$-ray emission we
calculate the emission from: Relativistic Bremsstrahlung, IC emission,
and Pion decay. Fig.~\ref{fig:gamma-ray_emission1} shows that IC
emission is dominant at low energies whilst Pion decay is dominant at
higher energies. For $\zeta_{\rm B}=\zeta_{\rm rel-e}=\zeta_{\rm
rel-p}=5\%$ the observed flux is considerably weaker than the
sensitivity of \textit{Fermi}, \textit{HESS}, \textit{HESS-II}, and
\textit{CTA}. In the case of efficient particle acceleration
($\zeta_{\rm rel-p}=50\%$) this gap is reduced and there is a
tentative possibility of a detection with \textit{CTA}
(Fig.~\ref{fig:gamma-ray_emission2}). These results then imply that
$\gamma$-ray emission will not be detectable from a single MYSO.

We close with the note that the possability of detecting non-thermal
$\gamma$-ray emission may be increased if consideration is given to
the cumulative effects of a cluster of massive stars, or, in the case
of efficient particle acceleration, from single MYSOs more massive
than our test case O8V star.

\acknowledgements ERP thanks the University of Leeds for funding
through a Henry Ellison Scholarship. JMP gratefully acknowledges
funding from the Royal Society. We would also like to thank Jim Hinton
for useful conversations.




\end{document}